# The Gaussian phenotype of biological measurements

Norichika OGATA

Nihon BioData Corp., Takatsu, Kawasaki, Kanagawa, 213-0033, JAPAN.
E-mail: norichik@nbiodata.com

## Abstract

Biological measurements are commonly assumed to approximate Gaussian distributions, and assessment of normality is routinely performed as a prerequisite for statistical analysis. However, whether the degree of Gaussianity itself contains biological information has remained largely unexplored. Here, we systematically quantified the Gaussianity of biological measurements using the root mean square error of normal quantile–quantile plots (QQ-RMSE). A reference distribution was constructed from 10,249 biological measurements obtained from the National Health and Nutrition Examination Survey (NHANES) 1999–2023. This reference enabled direct comparison of the Gaussian phenotype, defined as the degree to which a biological measurement approximates a Gaussian distribution. Biological measurements exhibited characteristic Gaussian phenotypes. Structural and capacity-related traits, including body measurements, grip strength, spirometry, and red blood cell count, consistently showed strong Gaussianity. Homeostatically regulated variables, such as total cholesterol, also exhibited low QQ-RMSE values. In contrast, biomarkers reflecting dynamic physiological responses or pathological processes, including triglycerides, C-reactive protein, liver enzymes, serum creatinine, and urinary albumin, showed progressively larger deviations from Gaussianity. Furthermore, biological normalization substantially improved Gaussianity: the albumin-to-creatinine ratio consistently exhibited lower QQ-RMSE values than urinary albumin alone across all NHANES survey cycles. These findings indicate that Gaussianity is not merely a statistical assumption but a measurable biological property. We propose the concept of the Gaussian phenotype, in which the degree of Gaussianity reflects fundamental biological mechanisms governing variability. This framework provides a quantitative reference for interpreting biological measurements and establishes Gaussianity as a biologically meaningful phenotype. This study establishes the first reference atlas of Gaussianity.

## 1. Introduction

Biological measurements occupy a central role in biology and medicine. Variables such as body size, blood cell counts, biochemical markers, and physiological functions are routinely analyzed using statistical methods that assume approximate Gaussian distributions [1,2]. Consequently, assessment of normality is generally treated as a prerequisite for selecting appropriate statistical tests rather than as a subject of biological investigation. However, biological measurements exhibit remarkably diverse distributional shapes. While some variables closely approximate Gaussian distributions, others are strongly skewed or heavy-tailed. Traditionally, such deviations have been regarded primarily as statistical problems that should be corrected by data transformation or addressed using non-parametric analyses [1,3,4]. Previous studies have primarily regarded Gaussianity as a prerequisite for statistical inference. In contrast, we treat Gaussianity itself as an observable biological property. Relatively little attention has been paid to the possibility that the degree of Gaussianity itself may reflect underlying biological mechanisms. The idea that Gaussianity may represent biological information was first suggested in our previous analysis of body-weight distributions in the beetle *Anomala albopilosa [5]*. That study indicated that deviations from Gaussianity could reflect biological organization rather than measurement error or sampling variability. Nevertheless, it remained unclear whether this observation represented a phenomenon specific to a single biological trait or a more general property shared across biological measurements. Addressing this question requires a comprehensive collection of heterogeneous biological variables measured under standardized conditions. The National Health and Nutrition Examination Survey (NHANES) provides an ideal resource because it contains thousands of independently measured variables spanning anthropometry, clinical chemistry, hematology, physiology, imaging, nutrition, and environmental exposure [6,7]. Such diversity enables systematic comparison of distributional properties across fundamentally different biological measurements. In the present study, we quantified the Gaussianity of 14,653 NHANES variables using the root mean square error of normal quantile–quantile plots (QQ-RMSE). Rather than treating Gaussianity as a statistical assumption, we considered it as a measurable characteristic of biological measurements. Using the resulting reference distribution, we examined how Gaussianity differs among biological measurements and whether systematic biological principles underlie these differences. We propose that the Gaussian phenotype constitutes a previously unrecognized phenotype of biological measurements, reflecting the mechanisms that generate biological variation rather than merely satisfying statistical assumptions.

## 2. Materials and Methods

### 2.1 Data source

All biological measurements analyzed in this study were obtained from the National Health and Nutrition Examination Survey (NHANES) 1999–2023, which follows a complex multistage probability sampling design described previously. NHANES is a nationally representative health survey of the United States population that includes anthropometric measurements, laboratory tests, physiological examinations, imaging data, environmental exposure assessments, and questionnaire-based variables. Publicly available datasets and data dictionaries were downloaded from the Centers for Disease Control and Prevention (CDC) NHANES website. Only numerical variables representing biological measurements were included in the present study. Administrative variables, participant identifiers, dates, sampling weights, and other non-biological variables were excluded.

### 2.2 Variable selection

Each numerical variable was evaluated independently. Variables were excluded if they contained fewer than three finite observations, had zero variance, or appeared to represent categorical variables. Categorical-like variables were operationally defined as variables having fewer unique values than the square root of the number of finite observations. Variables were analyzed separately for each NHANES cycle. Identical biological measurements obtained in different survey cycles were treated as independent variables when constructing the reference distribution. Because the objective of this study was to characterize the Gaussian phenotype of biological measurements rather than administrative metadata, variables describing examination procedures, time intervals, sampling weights, identifiers, or other operational information were excluded from biological interpretation.

### 2.3 Quantification of the Gaussian phenotype (QQ-RMSE)

The Gaussian phenotype of each biological variable was quantified using the root mean square error of a normal quantile–quantile plot (QQ-RMSE). For each variable, non-finite values, including missing values, NaN, and infinite values, were removed. Variables were excluded if they were non-numeric, contained fewer than three finite observations, had zero variance, or appeared to represent categorical variables. Categorical-like variables were operationally defined as variables for which the number of unique values was smaller than the square root of the number of finite observations. For each remaining variable, values were standardized to z-scores by subtracting the sample mean and dividing by the sample standard deviation. The standardized values were then sorted in ascending order. The corresponding theoretical quantiles were obtained from the standard normal distribution using plotting positions generated by the ppoints function in R. QQ-RMSE was calculated as

$$\text{QQ-RMSE} = \sqrt{\frac{1}{n}\sum_{i=1}^{n}\left(z_{(i)} - \Phi^{-1}(p_i)\right)^2}$$

where $z_{(i)}$ denotes the $i$-th ordered standardized value, $p_i$ denotes the corresponding plotting position generated by the R function ppoints, and $\Phi^{-1}$ denotes the inverse cumulative distribution function of the standard normal distribution.

Smaller QQ-RMSE values indicate that the observed standardized distribution more closely follows the theoretical standard normal distribution, whereas larger values indicate stronger deviations from Gaussianity. Because all variables were standardized before calculation, QQ-RMSE reflects the shape of the distribution rather than its original scale or units. Unlike conventional normality tests, QQ-RMSE provides a continuous measure of Gaussianity rather than a binary statistical decision. Normality tests such as the Shapiro–Wilk test are strongly influenced by sample size and are primarily designed for hypothesis testing. In contrast, QQ-RMSE directly quantifies the magnitude of deviation from the Gaussian distribution and therefore enables comparison of distributional shape across heterogeneous biological variables. In the present study, QQ-RMSE was not used as a normality test. Instead, it was used as a quantitative measure of the Gaussian phenotype, defined as the degree to which a biological measurement approximates a Gaussian distribution.

## 2.4 Reference distribution of the Gaussian phenotype

QQ-RMSE values were calculated for all eligible biological variables. The resulting distribution was used as a reference population for evaluating the Gaussian phenotype of individual biological measurements. Percentile values were calculated from the empirical distribution of QQ-RMSE. For practical interpretation, Gaussianity was classified into five grades (A–E) according to the reference percentiles.

## 2.5 Biological classification

Representative biological variables were selected from major NHANES categories, including body measurements, hematology, clinical chemistry, blood pressure, spirometry, dual-energy X-ray absorptiometry (DXA), and urinalysis, to illustrate the relationship between biological function and the Gaussian phenotype. For comparisons involving biologically related variables, identical measurements acquired in different NHANES cycles were analyzed independently to evaluate the reproducibility of the Gaussian phenotype.

## 2.6 Albumin-to-creatinine ratio

Urinary albumin-to-creatinine ratios (ACR) were calculated by merging urinary albumin and urinary creatinine datasets using the unique participant identifier (SEQN). Albumin concentrations (URXUMS) and urinary creatinine concentrations (URXUCR) obtained from the same participant were combined within each survey cycle. QQ-RMSE values of ACR were compared with those of urinary albumin and urinary creatinine to evaluate the effect of biological normalization on the Gaussian phenotype.

## 2.7 Software

All analyses were performed using R version 4.4.1 [8]. To download NHANES data, package nhanesA were used [9]. Custom R scripts developed for this study were used for calculation of QQ-RMSE and generation of the reference distribution. To facilitate practical application, a web-based calculator was developed that allows users to upload numerical data and automatically calculate QQ-RMSE, percentile rank, and Gaussianity grade relative to the reference distribution established in the present study [10,11]. BioNormality QQ-RMSE Reference is good at visualization and QQ-RMSE Reference is good at accurate ranking. The complete source code used for calculation of QQ-RMSE is provided as Supplementary Code 1.

```
## Supplementary Code 1
## Calculation of QQ-RMSE used in the present study

qq_rmse <- function(x) {

  # Exclude non-numeric variables
  if (!is.numeric(x))
    return(NA_real_)

  # Remove missing and non-finite values
  x <- x[is.finite(x)]

  # Require at least three observations
  if (length(x) < 3)
    return(NA_real_)

  # Exclude variables with no variation
  if (sd(x) == 0)
    return(NA_real_)

  # Exclude variables with fewer unique values than sqrt(n)
  if (length(unique(x)) < sqrt(length(x)))
    return(NA_real_)

  # Standardize observations (z-scores)
  z <- as.numeric(scale(x))
  z <- sort(z)

  # Calculate theoretical quantiles of the standard normal distribution
  q <- qnorm(ppoints(length(z)))

  # Calculate QQ-RMSE
  sqrt(mean((z - q)^2))
}
```

# 3. Results

## 3.1 Construction of the reference distribution of the Gaussian phenotype

A total of 14,653 variables were downloaded from the NHANES 1999–2023 datasets. Variables representing participant identifiers, sampling weights, questionnaire responses, imaging-derived measurements, laboratory quality-control variables, and other non-biological variables were excluded. The final reference dataset consisted of 10,249 biological measurements (Fig. 1A). Dataset is available at Zenodo [12]. QQ-RMSE was calculated independently for each biological variable. The resulting empirical distribution exhibited a wide range of Gaussian phenotypes, extending from variables that closely approximated a Gaussian distribution to variables showing substantial deviation from Gaussianity (Fig. 1B). The distribution was right-skewed, with most biological measurements exhibiting intermediate QQ-RMSE values. To facilitate interpretation, the empirical reference distribution was divided into five

Gaussianity grades according to percentile thresholds (Fig. 1C). Grade A represented the 5% of variables with the smallest QQ-RMSE values, whereas Grade E represented variables above the 80th percentile. The corresponding QQ-RMSE thresholds were 0.1219, 0.2866, 0.4816, and 0.8668. Representative examples of each Gaussianity grade are shown in Fig. 2.

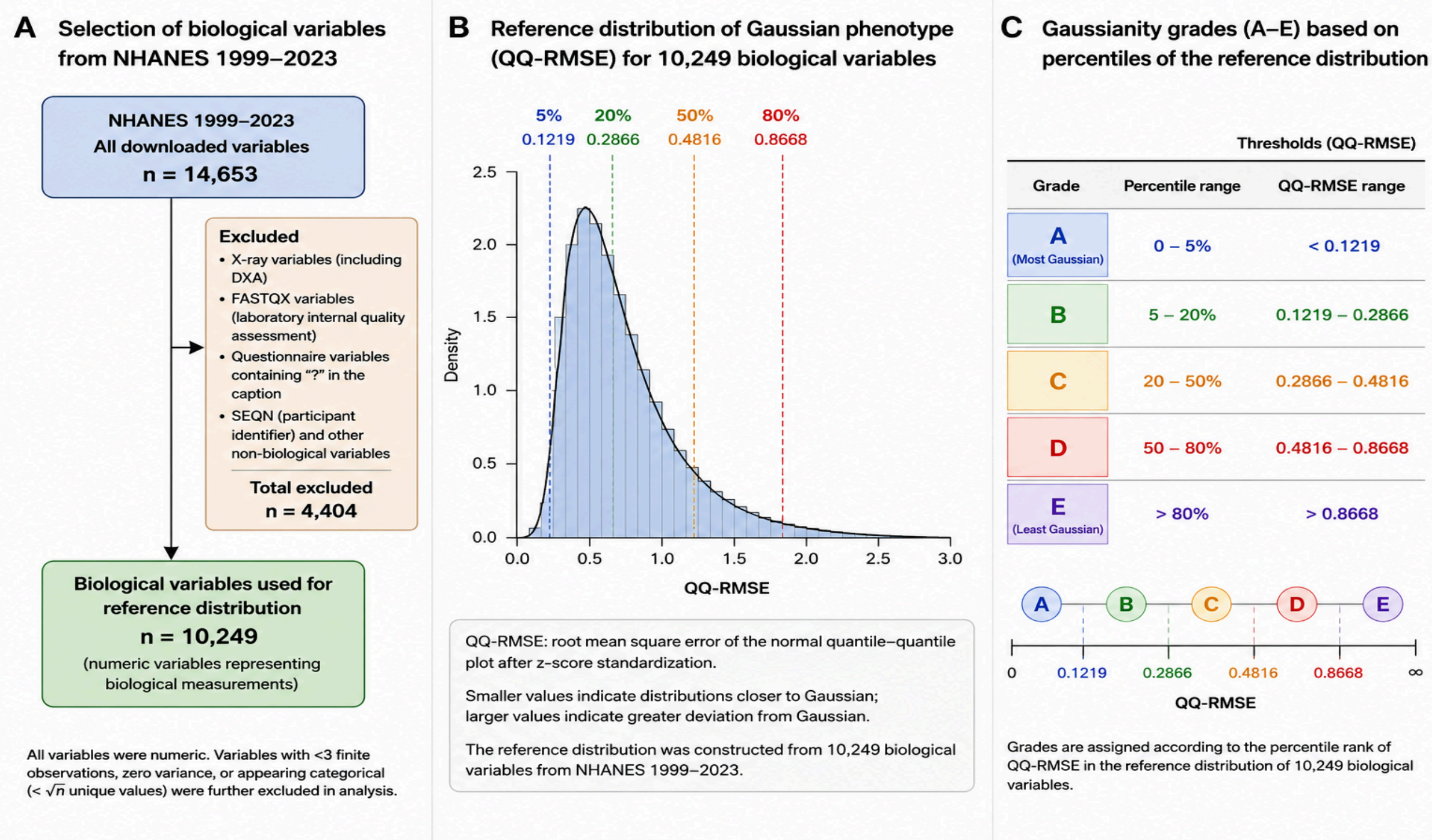

**Figure 1. Reference distribution of Gaussian phenotype overview.**

## 3.2 Representative Gaussian phenotypes

Representative variables from each Gaussianity grade are shown in Fig. 2. Variables assigned to Grade A exhibited distributions that closely approximated the theoretical Gaussian distribution, with nearly linear normal quantile–quantile plots. In contrast, variables assigned to higher grades showed progressively larger deviations from Gaussianity. Upper leg length was selected as a representative Grade A variable and exhibited an almost ideal Gaussian distribution (QQ-RMSE = 0.03). Serum total cholesterol represented Grade B (QQ-RMSE = 0.15), showing only modest deviation from Gaussianity. White blood cell count represented Grade C (QQ-RMSE = 0.36), exhibiting noticeable right-skewness. Triglyceride represented Grade D (QQ-RMSE = 0.82), characterized by a pronounced right tail. Serum total IgE antibody represented Grade E (QQ-RMSE = 0.92), displaying an extremely right-skewed distribution (Fig. 2). These representative variables demonstrate that biological measurements span a continuous spectrum of Gaussian phenotypes.

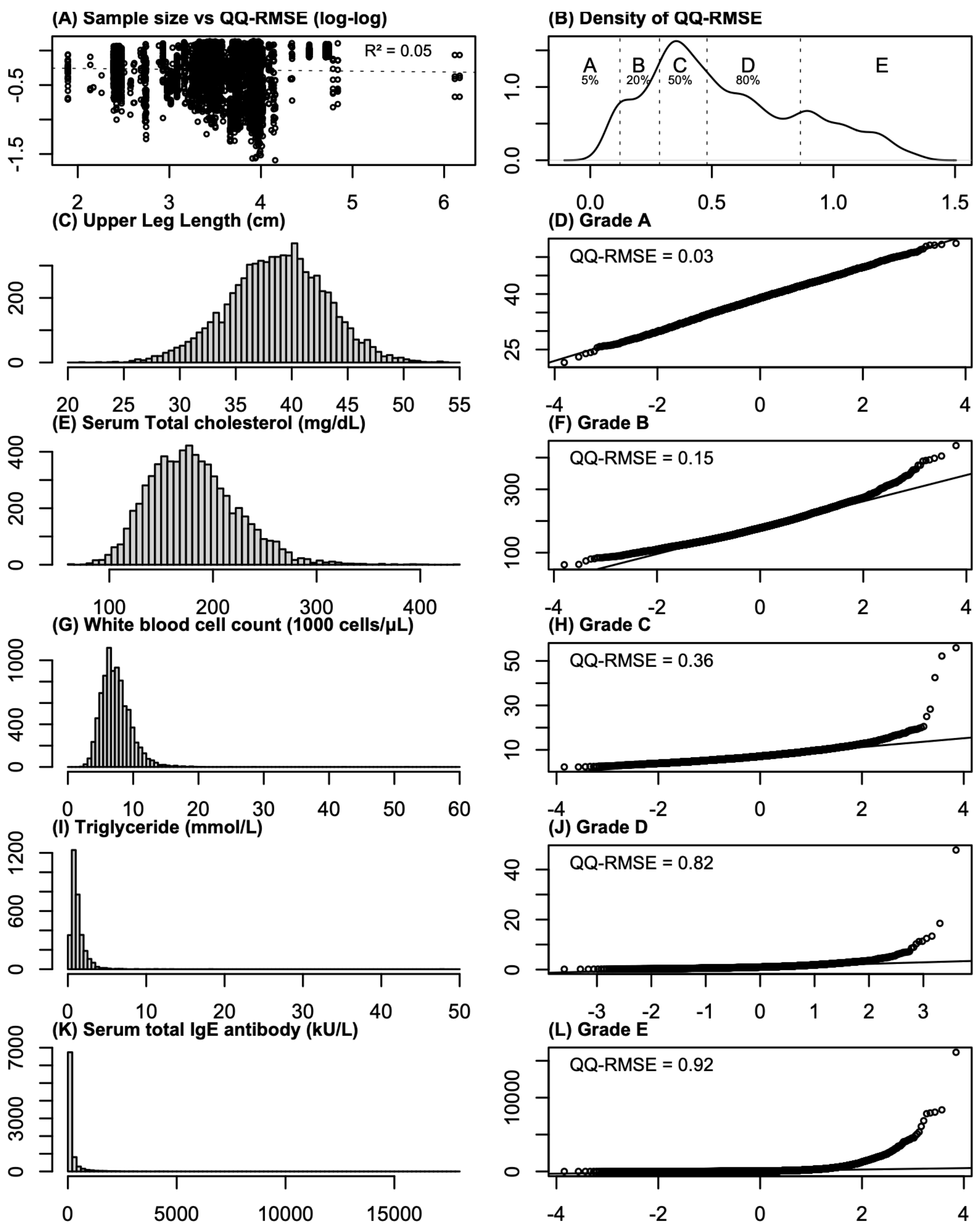

**Figure 2. Reference distribution of QQ-RMSE values across NHANES variables.**
(A) Relationship between sample size and QQ-RMSE. (B) Density distribution of QQ-RMSE with grade boundaries defined by the 5th, 20th, 50th, and 80th percentiles. (C–L) Representative histograms and corresponding Q-Q plots for Grades A–E. Panels C, E, G, I, and K show representative histograms for Grades A–E, and the adjacent panels (D, F, H, J, and L) show the corresponding normal Q-Q plots.

### 3.3 Biological measurements exhibit characteristic Gaussian phenotypes

QQ-RMSE values were reproducible across NHANES survey cycles for the same biological measurements, indicating that Gaussian phenotypes were characteristic properties of individual variables rather than random fluctuations. Distinct biological measurements consistently occupied different regions of the reference distribution (Fig. 3).

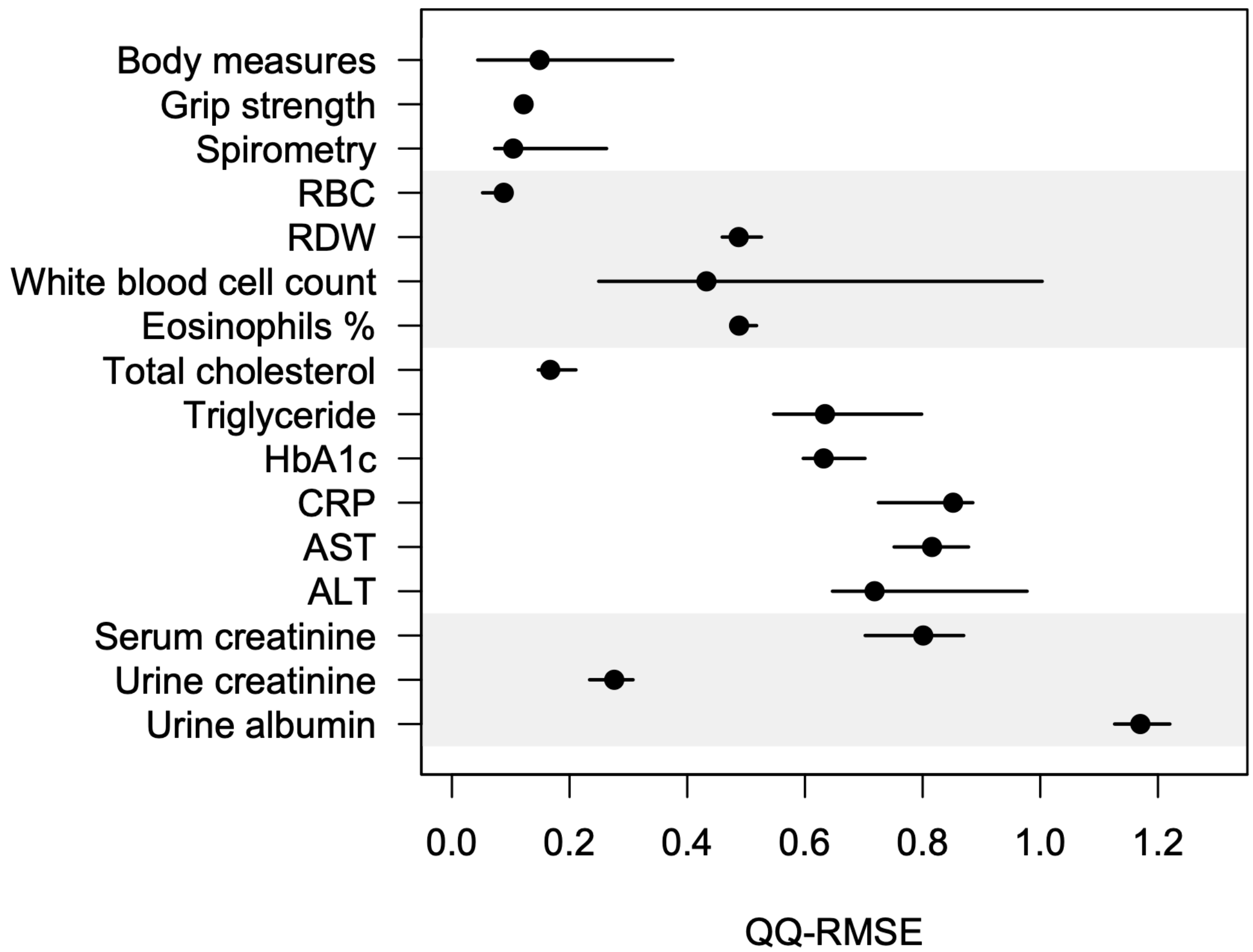

**Figure 3. Characteristic Gaussian phenotypes of representative biological measurements across NHANES survey cycles.**

QQ-RMSE values of representative biological measurements were calculated independently for each NHANES survey cycle. Horizontal lines indicate the range of QQ-RMSE values observed across survey cycles, and filled circles represent the median QQ-RMSE for each biological measurement. Structural and capacity-related traits, including body measurements, grip strength, spirometry, and red blood cell (RBC) count, consistently exhibited low QQ-RMSE values. Homeostatically regulated measurements, such as total cholesterol, also showed relatively small QQ-RMSE values. In contrast, biomarkers associated with inflammation, tissue injury, and renal dysfunction, including C-reactive protein (CRP), aspartate aminotransferase (AST), alanine aminotransferase (ALT), serum creatinine, and urinary albumin, consistently exhibited larger deviations from Gaussianity. The reproducibility of QQ-RMSE across independent survey cycles indicates that the Gaussian phenotype is a characteristic property of individual biological measurements.

Body measurements generally exhibited relatively small QQ-RMSE values. For example, upper leg length showed one of the smallest QQ-RMSE values among all variables, whereas head circumference and upper arm length exhibited only moderate deviations from Gaussianity. Similarly, grip strength and spirometric measurements consistently showed low QQ-RMSE values across survey cycles. Hematological measurements displayed substantial heterogeneity. Red blood cell count consistently exhibited very low QQ-RMSE values, whereas red cell distribution width (RDW) showed considerably larger values despite being obtained from the same hematological analysis. White blood cell count exhibited intermediate to large QQ-RMSE values, and leukocyte differential measurements, including eosinophil percentage and lymphocyte count, generally exhibited larger deviations from Gaussianity than red blood cell count.

Clinical chemistry measurements also exhibited characteristic Gaussian phenotypes. Total cholesterol consistently showed relatively low QQ-RMSE values, whereas triglycerides exhibited substantially larger values across all survey cycles. Glycohemoglobin (HbA1c) showed intermediate QQ-RMSE values. Biomarkers associated with inflammation or tissue injury, including C-reactive protein (CRP), aspartate aminotransferase (AST), and alanine aminotransferase (ALT), consistently exhibited large QQ-RMSE values. Renal biomarkers likewise differed in their Gaussian phenotypes. Serum creatinine exhibited intermediate to large QQ-RMSE values, whereas urinary creatinine showed considerably smaller values. Urinary albumin consistently exhibited among the largest QQ-RMSE values observed in routine clinical chemistry.

Notably, measurements obtained using the same analytical platform frequently exhibited markedly different QQ-RMSE values. For example, red blood cell count and red cell distribution width, as well as total cholesterol and triglycerides, consistently occupied different regions of the reference distribution despite being measured using identical laboratory methodologies. Collectively, these observations demonstrate that biological measurements possess characteristic Gaussian phenotypes that are reproducible across independent survey cycles and differ substantially among biological variables. Because urinary albumin exhibited one of the largest QQ-RMSE values among routinely measured clinical biomarkers, we next examined whether biologically meaningful normalization altered its Gaussian phenotype.

## 3.4 Biological normalization improves the Gaussian phenotype

To examine whether biologically meaningful normalization influences the Gaussian phenotype, the urinary albumin-to-creatinine ratio (ACR) was calculated by matching urinary albumin and urinary creatinine measurements using participant identifiers (SEQN). Urinary albumin consistently exhibited among the largest QQ-RMSE values of routinely measured clinical biomarkers, ranging from 1.12 to 1.22 across eight NHANES survey cycles (Fig. 4). In contrast, urinary creatinine exhibited substantially smaller QQ-RMSE values (0.25–0.31). Following normalization by urinary creatinine, the QQ-RMSE of ACR decreased to 0.42–0.87 in every survey cycle examined. The reduction in QQ-RMSE was consistently observed across all eight NHANES cycles, with improvements ranging from 0.25 to 0.75 relative to urinary albumin alone. Although ACR remained less Gaussian than urinary creatinine, normalization substantially reduced the deviation from Gaussianity while preserving the biological measurement. Thus, normalization by urinary creatinine consistently reduced QQ-RMSE across all NHANES survey cycles examined.

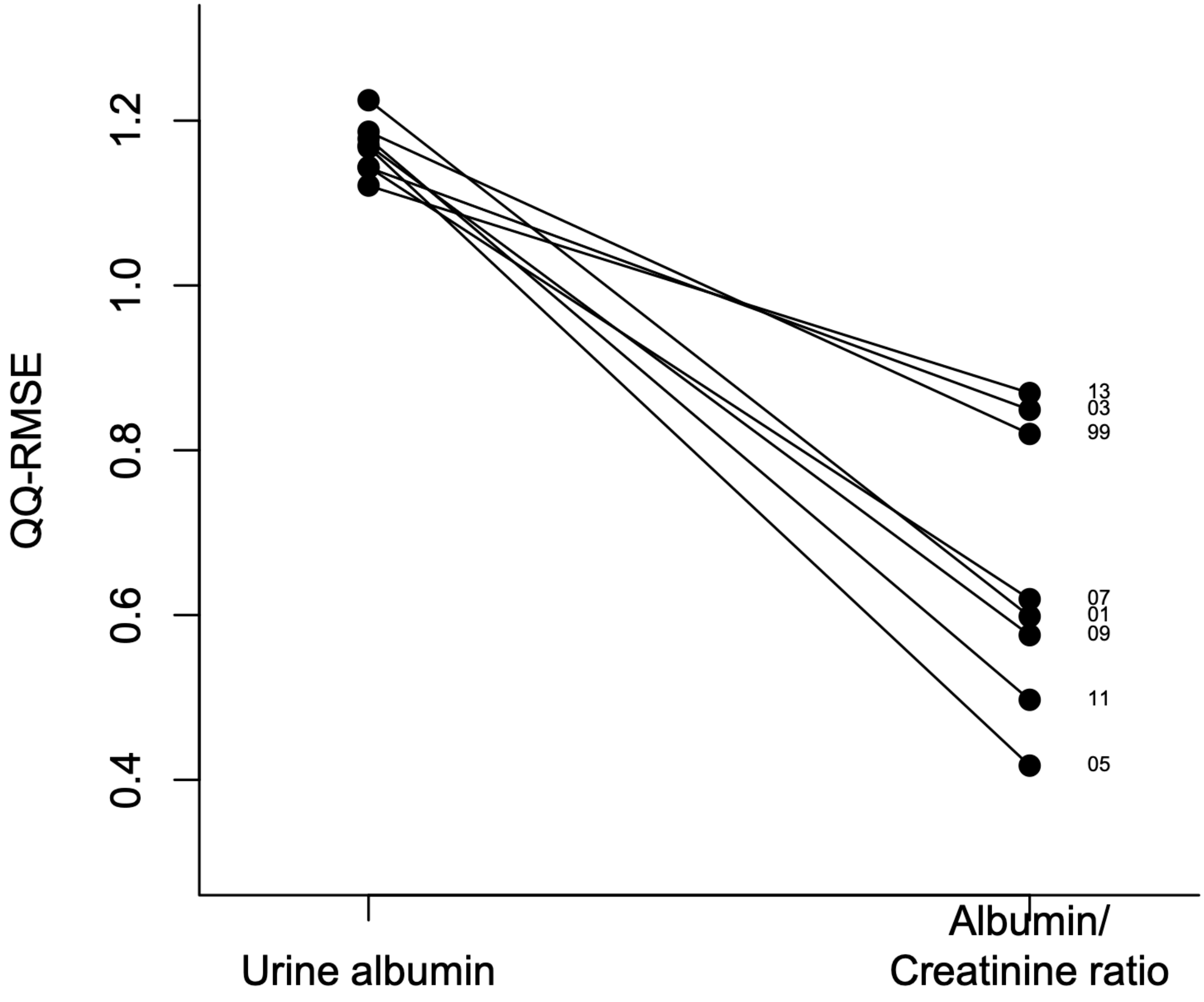

**Figure 4. Biological normalization improves the Gaussian phenotype of urinary albumin.**

QQ-RMSE values of urinary albumin and the urinary albumin-to-creatinine ratio (ACR) were compared across eight NHANES survey cycles. Each line connects measurements from the same survey cycle before and after normalization by urinary creatinine. Urinary albumin consistently exhibited large deviations from Gaussianity, whereas normalization by urinary creatinine reduced QQ-RMSE in every survey cycle. These results demonstrate that biologically meaningful normalization can substantially improve Gaussianity while preserving biological information.

## 4. Discussion

The Gaussian distribution has long served as the default statistical model for biological measurements. Consequently, departures from Gaussianity are generally regarded as statistical inconveniences that require transformation or the application of non-parametric methods. Our results suggest a fundamentally different interpretation. Analysis of 10,249 biological variables demonstrated that Gaussianity itself varied widely among measurements. QQ-RMSE values ranged from distributions that were nearly indistinguishable from an ideal Gaussian distribution to highly skewed distributions. These observations indicate that Gaussianity is not a universal property of biological measurements. Instead, the degree of Gaussianity represents a measurable characteristic of biological systems.

The concept that Gaussianity itself may contain biological information did not originate from the present study. Previous work on body weight distributions in the beetle *Anomala albopilosa* (QQ-RMSE 0.04862233) suggested that the degree of Gaussianity could reflect biological characteristics rather than statistical noise . The present study substantially extends this concept by demonstrating that Gaussianity varies systematically across thousands of independent biological measurements in humans. Rather than being restricted to a particular organism or trait, Gaussianity appears to represent a general property of biological measurements.

Based on these observations, we propose that the Gaussian phenotype of biological measurements is governed by four biological principles.

The first principle is <u>additivity</u>. Biological measurements determined by the combined effects of numerous independent factors tend to approach Gaussian distributions. Body dimensions, bone mineral density, grip strength, spirometric measurements, and red blood cell counts all exhibited relatively small QQ-RMSE values. These variables reflect cumulative contributions from genetics, development, nutrition, aging, and other largely independent biological processes. Their distributions are therefore consistent with expectations from the central limit theorem operating in biological systems.

The second principle is <u>homeostasis</u>. Some biological variables are actively stabilized by physiological feedback mechanisms. Total cholesterol and several other tightly regulated physiological measurements exhibited substantially greater Gaussianity than variables that respond rapidly to environmental or pathological stimuli. Homeostatic regulation constrains biological variation and limits extreme values, thereby maintaining relatively symmetric distributions despite continuous physiological perturbations.

The third principle is <u>unidirectional pathological response</u>. Unlike structural or homeostatically regulated traits, many biomarkers increase primarily in response to disease. C-reactive protein, aspartate aminotransferase, alanine aminotransferase, serum creatinine, and urinary albumin all exhibited substantially larger QQ-RMSE values. These measurements remain low in healthy individuals but increase dramatically during inflammation, tissue injury, or organ dysfunction, producing pronounced right-skewed distributions. A occiput-to-wall distance, a clinical screening measure for spinal deformity, which was zero-inflated because the vast majority of participants had a normal value (0 cm) (Figure 5). Thus, strong deviations from Gaussianity may reflect biological response mechanisms rather than measurement variability.

The fourth principle is <u>biological normalization</u>. Appropriate normalization can reduce biologically irrelevant variation and increase Gaussianity. A clear example was provided by the urinary albumin-to-creatinine ratio (ACR). Although urinary albumin alone consistently exhibited highly non-Gaussian distributions, normalization by urinary creatinine reduced QQ-RMSE in every NHANES cycle examined. This observation suggests that Gaussianity is sensitive not only to biological variability itself but also to the effectiveness of biologically meaningful normalization procedures.

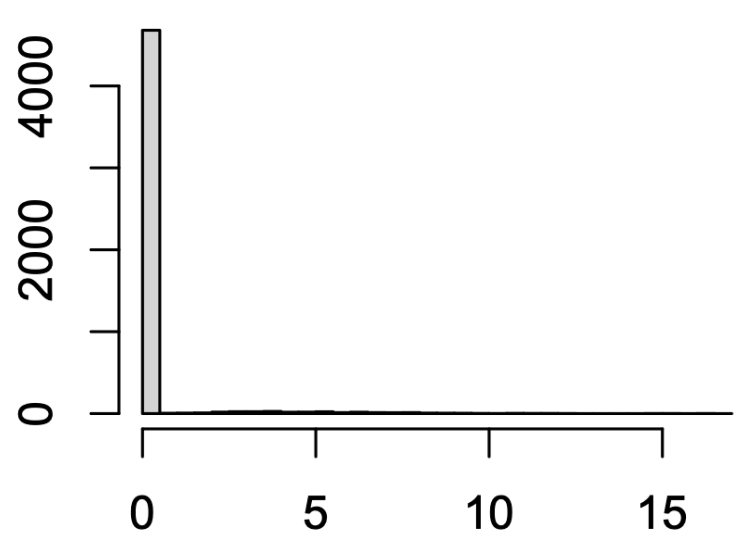

**Figure 5. Zero-inflated distribution of occiput-to-wall distance in the NHANES population.**

Histogram of occiput-to-wall distance measured in NHANES participants. The distribution is strongly zero-inflated because the vast majority of participants had a normal value of 0 cm, whereas only a small proportion exhibited non-zero values indicative of spinal deformity or postural abnormality. This variable represents a typical example in which a biologically meaningful floor effect produces a marked deviation from a Gaussian distribution (QQ-RMSE 1.03).

These four principles also explain why Gaussianity reflects biological function rather than measurement modality. Measurements obtained using the same analytical technology frequently exhibited markedly different Gaussian phenotypes. For example, total cholesterol and triglycerides are both serum lipid measurements but showed substantially different QQ-RMSE values, reflecting their distinct physiological regulation. Similarly, serum creatinine, urinary creatinine, urinary albumin, and the albumin-to-creatinine ratio exhibited strikingly different Gaussian phenotypes despite their close biochemical relationship. Likewise, red blood cell count and red cell distribution width, although derived from the same hematological analysis, occupied very different positions within the reference distribution. These observations indicate that Gaussianity primarily reflects biological meaning rather than analytical methodology.

An important outcome of the present study is the establishment of a reference framework for interpreting Gaussianity. By constructing a reference distribution from 10,249 biological variables, newly measured variables can now be expressed using percentile rankings and Gaussianity grades. This framework transforms QQ-RMSE from a descriptive statistic into an interpretable biological index, allowing direct comparison among diverse biological measurements. Such a reference may prove useful for evaluating biomarker characteristics, assessing normalization procedures, and comparing biological datasets generated by different technologies.

Unlike conventional normality tests, the present framework enables users to evaluate newly measured biological variables relative to a large empirical reference population. To facilitate practical use, we developed a web application that automatically calculates QQ-RMSE, percentile rank, and Gaussianity grade.

Several limitations should be acknowledged. The reference distribution was derived exclusively from NHANES and therefore primarily reflects measurements obtained in a large United States population. Although NHANES encompasses exceptionally diverse physiological and clinical variables, additional validation in other populations and organisms will be required. Furthermore, QQ-RMSE characterizes one aspect of distributional shape and does not capture every property of complex biological distributions, such as multimodality or temporal dynamics.

In conclusion, we propose that Gaussianity should no longer be regarded merely as a statistical assumption. Instead, it should be considered a measurable biological phenotype. The Gaussian phenotype provides a unified framework for understanding why different biological measurements exhibit distinct distributional characteristics and offers a quantitative approach for interpreting the biological organization of measurement variability.

## References

1. Holt J, Lumsden JH, Mullen K. On transforming biological data to Gaussian form. Can J Comp Med. 1980;44: 43–51.

2. Curran-Everett D. Explorations in statistics: the assumption of normality. Adv Physiol Educ. 2017;41: 449–453.

3. Habibzadeh F. Data distribution: Normal or abnormal? J Korean Med Sci. 2024;39: e35.

4. Ghasemi A, Zahediasl S. Normality tests for statistical analysis: a guide for non-statisticians. Int J Endocrinol Metab. 2012;10: 486–489.

5. Ogata NN, Ogata N. An Accurate Bell Curve Discovered in Intraspecific Body Mass Distribution of 1k Beetles. 1 Oct 2025 [cited 9 Jul 2026]. Available: https://vixra.org/abs/2510.0143

6. Curtin LR, Mohadjer LK, Dohrmann SM, Montaquila JM, Kruszan-Moran D, Mirel LB, et al. The National Health and Nutrition Examination Survey: Sample Design, 1999-2006. Vital Health Stat 2. 2012; 1–39.

7. Zipf G, Chiappa M, Porter KS, Ostchega Y, Lewis BG, Dostal J. National health and nutrition examination survey: plan and operations, 1999-2010. Vital Health Stat 1. 2013; 1–37.

8. R Core Team. R: A Language and Environment for Statistical Computing. 2024. Available: https://www.R-project.org/

9. Ale L, Gentleman R, Sonmez TF, Sarkar D, Endres C. nhanesA: achieving transparency and reproducibility in NHANES research. Database (Oxford). 2024;2024: baae028.

10. BioNormality QQ-RMSE Reference. [cited 9 Jul 2026]. Available: https://nbiodata.com/qqrmse/qqrmse_reference_tool.html

11. QQ-RMSE Reference. [cited 9 Jul 2026]. Available: https://nbiodata.com/qqrmse/qqbio.html

12. Ogata N. QQ-RMSE values and sample sizes of 10,249 NHANES variables. Zenodo; 2026. doi:10.5281/ZENODO.21284307